\documentstyle[12pt]{article}

\def\be{\nopagebreak[3]\begin{equation}}
\newcommand{\ee}{\end{equation}}
\def\ba{\begin{array}}
\def\ea{\end{array}}

\def\inf{\infty}
\def\th{\theta}
\def\D{\Delta}
\def\pa{\partial}

\newcommand{\tr}{{\rm tr}\,}

\renewcommand{\d}{\partial}

\newcommand{\la}{\lambda}
\newcommand{\ra}{\rightarrow}
\newcommand{\khi}{\chi}
\newcommand{\te}{\theta}
\newcommand{\om}{\omega}

\begin{document}
\begin{titlepage}
\begin{flushright}
February, 1995
\end{flushright}

\bigskip

\begin{center}
{\LARGE
2D Principal Chiral Field at Large N as a Possible
Solvable 2D String Theory}

\vskip 0.7truecm
{\bf Vladimir A. Kazakov \\}

 \vskip 0.3 cm
{\it Laboratoire de Physique Th\'eorique de l'Ecole Normale Sup\'erieure%
\footnote{
Unit\'e Propre du Centre National de la Recherche
Scientifique, associ\'ee \`a l'\'Ecole Normale Sup\'erieure et \`a
l'Universit\'e de Paris-Sud.}
\\
24 rue Lhomond, 75231 Paris Cedex 05, France\\}
\vspace{1pc}

(based on the papers \cite{FKW} written
in collaboration with V.~A.~Fateev and P.~B.~Wiegmann)

\vspace{1pc}

\vskip 1.1in

{\large \bf Abstract}
\end{center}

We present the exact and explicit solution of the principal chiral field in
two dimensions for an infinitely large rank group manifold.
The energy
of the ground state  is explicitly
found for the external Noether's fields of an arbitrary magnitude.
At small field we found an inverse logarithmic singularity
in the ground state energy at the mass gap which indicates that at
$N=\infty$ the spectrum of the theory contains extended objects rather than
 pointlike particles.

\vfill
\end{titlepage}

  \baselineskip19pt plus 2pt minus 1pt

\section{Introduction}

The progress in the modern quantum field theory, especially in the
solution of realistic problems like QCD, quantum gravity and strings
in physical dimensions, seems to be considerably slown down because of
a very limited list of exactly solvable models of many interacting
degrees of freedom. Currently, this list includes the following
relatively well explored systems:

1. Quantum field theories and statistical mechanical spin models for
the dimensions $D \le 2$;

2. Strings or 2d gravity coupled to matter with central charge $c_{m}\le1$;

3. Quantum gravity for $d\le2$.

All the attempts to make some substantial progress on the way to
realistic dimensions are usually confronted to the obstacles which
every time seem to have a very similar mathematical origin. It is
clear that some principally new mathematical tools will be needed to
attack in the future all these problems.  In the absens of them we may
try to use the old tools to get an insight to the physics of realistic
dimensions for the string theory.

One conceivable loophole would be to take some integrable two
dimensional field theory of $NxN$ matrix valued field. Its planar
diagrams might describe fluctuating world sheets of a string
propagating in the 1+1 dimensional space.

Essentially, there exists only one candidate - the principal chiral
field (PCF) on, say, $SU(N)$ group manifold. It can be defined by the
following action:

\be
S = {N
\over 2 \la_0 } \int d^2x \  \tr [ \d_\mu g^{\dag} \d_\mu g  ]
\label{action}
\ee
 where $g$ is an $N$x$N$ unitary matrix. Its large $N$ solution has
 been anticipated for a long time to follow from its finite $N$
 solution
\cite{PolWieg,W1,W2}. It was finally explicitly established in our
paper
\cite{FKW}. We will follow the guidelines of this solution.

The partition function of the model reads as:
\be
Z=\int [D g(x)]_H \exp S
\label{parti}
\ee
where the integration over $g$ goes with the Haar measure for $SU(N)$
at every point of the 2D space.  To test physics in this
asymptotically free theory we have to introduce somehow an energy
scale. Technically the simplest way to do it is to take the theory on
a cylinder with the space compactified on the space interval $[0,L]$
and to introduce the following twist of the field $g$:
\be
g(x,t) \rightarrow e^{H_Lx}g(x,t)e^{H_Rx}
\label{twist}
\ee
where  $H_{R(L)}= diag(h_1, h_2-h_1, ...,
h_{N-1}-h_{N-2}, -h_{N-1})$ is an element of the Cartan subalgebra.

It amounts to introducing in eq.(\ref{action}) the covariant derivative
\be
D_\mu g = \d_\mu g -i  \delta_{\mu0} (H_L g+g H_R)/2
\label{cov}
\ee
instead of the usual derivative $\d_\mu g$. In what follows, we shall
consider only the case $H_L=H_R=H$.

In the hamiltonian language one adds to the hamiltonian of the PCF a term
$tr (H_L Q_L+H_R Q_R)/2$
to the hamiltonian of the theory corresponding to the lagrangian (1).
Here
$Q_L= \int d^2x g \d_0 g^{-1} $ and
$Q_R= \int d^2x g^{-1} \d_0 g $ are
the Noether's left and right charges.

The model is completely integrable for any N by means of the Bethe
ansatz approach \cite{W1} or the bootstrap procedure for the physical
S-matrix. The results show the presence of $N-1$ physical particles in
the spectrum of the theory obeying the masses:
\be
 m_l= m  {\sin({\pi \over N} l ) \over \sin({\pi \over N}  )  }
\label{mk}
\ee
where $l=1,...,N-1$ is the rank of a fundamental representation and
 $m=m_1$ is the mass of the vector particle.
In the two-loop approximation
it is
\be
m=  \Lambda { 1 \over \sqrt{\la_0} } exp({-4 \pi \over \la_0})
\label{mass}
\ee
where  $\Lambda$ is a cutoff.  These results are
in complete agreement with the renormalization group predictions for
the asymptotically free theory.

All particles are bound states of the vector particles.
They form the multiplets of the fundamental representations of $SU(N)$
algebra (antisymmetric tensors corresponding to columns of the Young
tableau). It follows from the last statement that $h_k$ is precisely
the chemical potential of the k-th type of the physical particles.

To demonstrate this last statement let us turn for a moment to the
one-dimensional PCF and calculate iths partition function in the
periodic time $(0,T)$ with the same twisted boundary conditions.

 The functional integral representation of the
partition function is
\be
Z(T,H_L,H_R)=\int[D g(t)]_H
e^{Ntr\int_0^T dt\pa_t  g^{-1}\pa_t  g},
\label{chiralZ}
\ee
where $[{\cal D}g(t)]_H$ is the Haar measure on the unitary group $U(N)$,
and the holonomies at the end points of the time interval are
$g(0)=e^{iH_L}$  and $g(T)=e^{iH_R}$ (with $e^{iH_L}$ and
$e^{iH_R}$ the eigenvalues of $g(0)$ and $g(T)$).
Due to the invariance of the Haar
measure at each moment of time, we have, after the introduction of a
new variable (connection) $A(t)=i g^{-1}\pa_t  g(t)$ the
following representation of $Z(T,H_L,H_R)$:
\be
Z(T,H_0,H_T)=\int D A(t)e^{-Ntr\int_0^Tdt A(t)^2}
\,\,\delta\Bigl(\bigl[Te^{i\int_0^TdtA(t)}\bigr]e^{iH_L},e^{iH_R}\Bigr).
\label{chiral2Z}
\ee
Using the character expansion of the
group $\delta$-function:
\be
\delta(U,U')=\sum_R \chi_R(U) \chi_R(U')^*
\label{delt}
\ee
 and
the fact that
the character is the trace of the matrix element in a given
representation R:
$\chi_R(T\exp(i\int Adt))=tr_R[T\exp(i\int A^a\tau^a_Rdt)]$
($\tau^a_R$
is the $a$th generator of $U(N)$ in the $R$th irreducible representation),
we arrive after a simple gaussian integration in $A(t)$ (independent at
each moment $t$) at the conclusion that
\be
Z(T,H_L,H_R)=\sum_R e^{-T C_2(R)}\chi_R(e^{iH_L})\chi_R(e^{iH_R})
\label{heatk}
\ee
where $C_2(R)$ is the second Casimir of the group $U(N)$ in the
representation $R$.
This is also equivalent to the partition function of two dimensional
QCD on the cylinder with the
boundary holonomies specified at either end of the cylinder by the two
distributions $e^{iH_L}$ and $e^{iH_R}$.

Let us now recall the Weil formula for the character:
\be
\chi_R(e^{iH})={\det_{k,j}(e^{i(n_k-k+N)H_j}) \over \D(e^{iH})},
\label{ndefns}
\ee
 where $\D(x)=\prod_{i<j}(x_i-x_j)$ is the Van-der-Monde determinant.

Let us consider one of two characters entering the expression
(\ref{heatk}).  Due to the antisymmetry of the characters in $n_k$'s
we can retain in one of them only a diagonal term from the Weil
determinant under the sum over irreps (in other words, over integers
$n_k$):

\be
\exp(i\sum_k H_k n_k)= \exp(i\sum_k (h_k-h_{k-1})n_k) =\exp(-i\sum_k
h_k(n_{k+1}-n_k) )
\label{polyv}
\ee

Since $n_{k+1}-n_k$ is the number of columns of length $k$ in the
Young tableau of the irrep R (corresponding to the antisymmetric group
of k indices from the tensor describing the whole irrep R), we
conclude that $h_k$ is the chemical potential of k-vectors - the k-th
fundamental representations which constitute the whole irrep R. In the
2D PCF these k-vectors will correspond to the physical particles of
the type k.

\section{ Ground state energy and beta-function of the PCF at large N}

Let us turn back to two the two-dimensional PCF, formulate our main
result and then discuss it in details.

Let us make a special choice of $h_k$'s which will be technically the
simples one and quite suitable for the limit $N = \infty$:

\be
h_k= h { \sin({\pi \over N} k ) \over \sin({\pi \over N}  )  }
\label{hk}
\ee

Since every k-th type of particles is excited only if $h_k$ exceeds
$m_k$, we will have for this choice of $h_k$'s no particles excited
when $h \le m$, and all particles excited on equal footing when $h >
m$.

 We  show that the energy of the ground state is expressed in terms of
modified Bessel
functions:
\be
 f(h) \equiv {1\over N^2}
\Big({\cal E}(h)-{\cal E}(0)\Big) = - {h^2 \over 8\pi} B^2 I_1(B) K_1(B)
\label{fh}
\ee
where the parameter $B$ is defined through
\be
{m \over h} = B K_1(B)
\label{mh}
\ee

The distribution of rapidities of physical particles will obey the
simple semi-circle law with the support $B$. The parameter $B$ defines
the value of rapidity corresponding to the Fermi momentum of the fused
particles.  We shall see, that $B$ gives the most natural definition
of the renormalized (running) coupling constant :
\be
 {\bar \la}(h)= {4\pi \over
B}
\label{la}
\ee

With the definition (\ref{la})
of the running charge one can find from  eq. (\ref{mh}) the exact
beta-function:
\be
\beta({\bar \la}) = h {\d \over \d h}{\bar \la} = - {4\pi \over B^2}
{\d B \over \d \ln {h/m} } = - 4 \pi { K_1(B)\over B^2 K_0(B)}
\label{beteq}
\ee
or
\be
\beta({\bar \la}) =
- {1 \over 4\pi} {\bar \la}^2
 { K_1({4\pi \over {\bar \la}})\over K_0({4\pi \over {\bar \la}})}
=- {1 \over 4\pi} {\bar \la}^2
 \sum_{n=2}^{\infty} b_n \Big({{ \bar \la} \over 32 \pi}\Big)^n
\label{betaex}
\ee
where
\be
\ba{rcl}
&&  b_0=1, \ \ b_1=4, \ \ b_2=-8, b_3=64, \ \ b_4= -5^2\cdot 2^5,  \\
&&  b_5=13\cdot 2^{10}, \ \  b_6=-1073 \cdot 2^8, \ \ ,b_7=103 \cdot 2^{16}
, \ \ ...
\ea
\label{coefb}
\ee

\be
b_n \sim  -(-1)^n \sqrt{8/(\pi n)} (4n/e)^n, \ \ \ n \ra \infty
\label{coefbas}
\ee

\section{Physical consequences of the exact solution:week and strong
coupling limits}

Let us first look for the week coupling regime and find the link with
the results of the standard renormalized perturbation theory.
In the asymptotically free theory week coupling means the presence of
a big energy scale, namely, big $h/m$.
 From eq. (\ref{mh}) we conclude that it corresponds to $B \ra \infty$.
 It follows from the large $B$ asymptotics  of McDonald function

\be
{h \over m} =\sqrt{{2 \over \pi}} {e^B \over \sqrt{B}} \Big(1-{3 \over 8 B}
+ O(1/B^2)  \Big)
\label{hmB}
\ee
Solving
 for $B$ we obtain:
\be
B = \ln{h \over m} + {1 \over 2} \ln \ln {h \over m} + {1 \over 2}
\ln{\pi \over 2} + O({1 \over \ln {h \over m}})
\label{Bhm}
\ee
Using the large $B$ asymptotics  $I_1(B)$
we finally
find from (\ref{fh}) and
(\ref{Bhm}):
\be
16 \pi  f(h) = - h^2  B + O({h^2 \over B}) =
- h^2
\Big( \ln{h \over m} + {1 \over 2} \ln \ln {h \over m} + {1 \over 2}
\ln{\pi \over 2} + O({1 \over \ln {h \over m}}) \Big)
\label{bighf}
\ee
This result reproduces correctly one- and two-loop terms of the
perturbation theory as well as
 the universal non-perturbative constant ${1 \over 2} \ln{\pi
\over 2}$. We also see from the comparison of eq.(\ref{Bhm}) with the
standart 2-loop perturbation theory that our definition (\ref{la}) of
the running coupling was justified.

More than that: we can use the known formulae for the $1/B$
expansion of the product $I_1(B)K_1(B)$ to get the explicite
coefficients of the renormalized week coupling expansion:
\be
 f(h)/h^2 = -{1 \over 4 {\bar \la} } \Big(1-
\sum_{n=1}^{\infty} C_{2n} \Big({{\bar \la} \over 4\pi}\Big)^{2n}\Big)
\label{perex}
\ee
where
\be
C_{2n} = {2n+1 \over 2n-1 } {((2n)!)^3 \over (n!)^4 8^{2n}}
\ra_{n \ra \infty} 2/\sqrt{\pi n} ({ n \over  e})^{2n}
\label{coefs}
\ee
Note that the expansion goes only in even powers.

As we see, in spite of the fact that every coefficient represents a sum
over renormalized planar graphs, it grows factorially with the
order. Most probably this happens because of the renormalons (some
subsequence of logarithmically divergent graphs) giving the main
factorial contribution in each order noticed long time ago by 'tHooft.
 This means that we have an exponential number of graphs in
each order but some of them give $(2n)!$ contribution after the
momenta integration.  More than that: the series is a non-signchanging
one and thus non-Borel summable. Nevertheless, the free energy
perfectly exists for any finite ${\bar \lambda}$.  These phenomena
seem to be imminent for any asymptotically free field theory. It sheds
some doubts on the possibility to interpret the standard planar Feynman
diagrams in this theory as the fluctuating world sheets of some
string. The point is that in all examples of the solvable theories of
non-critical strings with the central charge of the matter $c \le 1$
the expansions in the cosmological constant (conjugated to the
invariant area of the world sheet, or, in the discretized version of
the string theory, to the number of vertices of the corresponding
random (Feynman) graph) have usually a finite radius of convergency,
which means that near the critical value of the cosmological coupling
the graphs are big. In the present case we have a zero radius of
convergency with respect to the renormalized coupling (because of
renormalons), which prevents the usual procedure of the
thermodynamical limit of a big world sheet.

The hope to interpret the model in terms of a new string theory gets
revived if one considers an opposite, nonperturbative limit of
small energies (of the order of the mass gap): $\Delta=h/m-1\ra 0$). This
corresponds
to $B \ra 0$.

At small $B$ asymptotics of the Bessel functions give a singular
behaviour on the threshold:
\be
 f(h)  \simeq - (m/2\pi)^2 {\Delta \over |\ln \Delta|},
 \ \ \Delta \ra 0
\label{freD}
\ee

It differs drastically from the threshold behaviour for a finite N theory
of massive  particles, where we would have  3/2 law (see e.g. \cite{W1}):
\be
 f_N(h) \sim  -m^2 (\Delta)^{3/2}
\label{freN}
\ee

The reason for that is that we fixed the mass scale at $m_{N/2}$ at
large N, and therefore the mass spectrum became continuous (we have
N-1 different masses on the finite interval). The simpler limit would
correspond to fixing $m_1$. In this case $m_k=k m_1$ and we would not
have any bound states.

It is interesting to compare the result  (\ref{freD})
 with  the $c=1$ matrix
 model \cite{Kaz} where the ground state energy behaves in a similar
way with respect to the cosmological constant $\la$:

\be
 E(\la)  \simeq  {(\la_c-\la)^2 \over |\ln (\la_c-\la)|},
 \ \ \Delta \ra 0
\label{c1}
\ee

The mechanism by which the inverse logarithmic behaviour with
respect to the cosmological constant occurs also requires a
parametrization through the fermi level of the corresponding fermions
(whose coordinates are the eigenvalues of a hermitean matrix field).
The Fermi level plays the role of a "hidden" parameter of the problem
and the eigenvalues give rise to an extra (Liouville) degree of
freedom of the theory.

\section{ Sketch of the solution}
Let us now sketch out the major steps of the solution of the PCF.

The easiest way to start with is the construction of the physical
S-matrix for the (vector$\times$vector) particles \cite{W2,Ab}. Following
the general guidelines of the bootstrap method one arrives to the
result :

 ${\cal S}=X(\theta) S(\theta) \otimes S(\theta)$.  Here $\theta$ is
the rapidity of a massive relativistic particle
($p^0=m\cosh\theta$,\,$p^1=m\sinh\theta$) and $X(\theta)$ is the
CDD-ambiguity factor which cannot be determined by the factorization,
unitary and crossing symmetry conditions. The $SU(N)$ unitary,
crossing invariant, factorized $S$-matrix of vector particles is well
known. It is
\be
 S(\theta)=u(\theta)(P^{+}+{\theta +i2\pi/N\over \theta-i2\pi/N}P^{-})
\label{S}
\ee
 where $P^{\pm}$ is  the projection operator onto symmetric
(antisymmetric) states.

\be
 u(\theta)={{\Gamma
(1-{{\theta}\over{2\pi i}})\Gamma ({{1}\over{N}}+{{\theta}\over{2\pi
i}})}\over{\Gamma (1+{{\theta}\over{2\pi i}})\Gamma
({{1}\over{N}}-{{\theta}\over{2\pi i}})}}
 \label{u}
 \ee

Finally, the CDD factor $X$ is chosen
to cancel all double zeros and double poles on the physical sheet
$0<Im\theta<\pi$:
\be
X(\theta)={\sinh({\theta\over 2}+{i\pi\over N})
\over \sinh({\theta\over 2}-{i\pi\over N})}
\label{S2}
\ee
This is the $S$-matrix of the vector particles. It has a pole on the
physical sheet at $\theta_b=2\pi i/N$ in the antisymmetric channel. It
corresponds to the first bound state ( the second rank antisymmetric
tensor) with a mass $m_2=m\sin(2\pi /N)/\sin(\pi/N)$. The $S$-matrix
of these particles can be also found by tensoring the vector
$S$-matrix (the fusion procedure). It also has a pole in the
antisymmetric channel, and so on. In this way the whole mass spectrum
(\ref{mk}) can be generated.

To find the thermodynamical properties of the ground state we have to
use the Bethe ansatz procedure. The idea is to put many vector
particles into a periodic box of the size $L$. In the thermodynamical
limit they will self-organize into higher bound states, and to find
the ground state energy we have to calculate the densities of physical
particles $\rho_k(\theta)$ for every type $k=1,2,...,N-1$.

The next step (which can be justified within the thorough
consideration of the Bethe procedure) consists from throwing away the
isotopic index structure of the S-matrix and using its scalar part
$S(\th)= u^2(\theta)X(\theta)$
 to write down the phase ballance eq. for a particle
moving around the box:
\be
\exp(imL\sinh \theta_{\alpha})=\prod^{\cal
N}_{\beta=1,\alpha\ne\beta} exp(i\phi(\theta_{\alpha}-\theta_{\beta}))
\label{ba1}
\ee
 where $exp(i\phi(\theta)) =u^2(\theta)X(\theta)$ .

To restore the whole $SU(N)\times SU(N)$ structure of the physical
states of the model we have to consider the solutions of this eq. for
complex values of $\th$'s. It is known that in the large $L$ limit the
vector particles form the ``strings'' of complex rapidities:
$\theta^{r,(l)}\rightarrow\theta^{(l)}+2r\pi i/N$, where $\theta^l$ is
a rapidity of the $l$-th particle and $r$ is an integer running
between $-l/2$ and $l/2$.  Substituting this into the Eq.(\ref{ba1})
and multiplying equations over $r$ we shall obtain the equations for
the rapidities of the state which contains ${\cal N}_l$ particles of
the kind $l$.  Taking the logarithm of both sides the Eq.(\ref{ba1})
we obtain
\be
Lm_l\sinh \theta^{(l)}_{\alpha}=2\pi J^{(l)}_{\alpha}+
\sum _{n=1}^{N-1}\sum^{{\cal
N}_n}_{\alpha=1,\ne\beta}\phi_{ln}(\theta_{\alpha}^{(l)}
-\theta_{\beta}^{(n)})
\label{ba3}
\ee
where $\phi_{ln}(\theta)=
\sum_{|r|<l/2,|r^{\prime}|<n/2}\phi(\theta+2ri\pi/N+2r^{\prime}i\pi/N)$
 is the
scattering phase of the $l$-th and the $n$-th particles, and integers
$J$ are the quantum numbers of the states. The energy of this state is
obviously \be E={1 \over L} \sum_{l=1}^{N-1} m_l
\sum_{\alpha=1}^{{\cal N}_l} \cosh\te_\alpha^{(l)}
\label{energy}
\ee

One can note here that the ground state of the system can be described
by a big Young tableau where every column of a length $k$ corresponds
to a physical particle of the type $k$ which transforms as k-th
fundamental representation of $SU(N)$.

 The next step is to find rapidities to minimize the energy
(\ref{energy}) in the thermodynamic limit ${\cal N}_l/L=n_l$, while
$L\rightarrow \inf$. We assume that in the ground state $\te$'s are
distributed smoothly between $-B_l$ and $B_l$ with a distribution
functions $\rho_l(\te)$ Than eq.(\ref{ba3}) implies the spectral
equations
\be
 {1\over 2\pi} m_l\cosh
\theta=
\sum_n\int^{B_l}_{-B_l}  R_{ln}(\te-\te')\rho_n(\te')d\te'
\label{ro}
\ee
 where $R_{ln}(\theta)=\delta_{ln}-{1\over 2\pi} {d\phi_{ln}(\theta) /
d\theta}$. The Fermi rapidities $B_l$ are determined by the number of
particles in the $l$ th representation: $ \int^{B_l}_{-B_l}
\rho_l(\te)d\te= n_l$.  The energy of the state is then
\be
 E=\sum_l\int ^{B_l}_{-B_l} m_l\cosh\te\rho_l(\te) d\te
\label{energy1}
\ee
An explicit form of the scattering kernel $R_{ln}$ was found in
\cite{W1}. Its Fourier transformation is
\be
R_{ln}(\om)=2
{{\sinh\Big(\pi\om(1-{l\over N})\Big)
\sinh\Big({\pi\om n\over N}\Big)}\over \sinh\pi\om }
\label{Four}
\ee
at $l>n$ and $R_{ln}=R_{nl}$.

Now we pass to the large N solution (found in \cite{FKW}) of this
rather complicated system of integral equations.

At large $N$ we can consider a particular distribution of fields $h_l$
which creates all different particles on equal footing, namely one
which follows the spectrum of masses (\ref{mk}): $ h_l=(h/m)m_l$
. This field creates ${\cal N}_l={\cal N} (m_l/m)$ particles in the
$l$-th representation (the most representative Young tableau).  In
this case all Fermi momenta are equal: $B_l=B$ and $\rho_l={1\over N}
(m_l/m)\rho$.  Then the spectral equations (\ref{ro}) can be easely
diagonalized. They reflect the structure of the Cartan matrix and
moreover have the same eigenvectors: $$
\sum_{l,n=1}^{N-1}\khi^{(p)}_l R_{ln}(\om)  \khi^{(p')}_n=
  R^{(p)}(\om)\delta^{p,p'}$$
\be
R^{(p)}(\om)={2 N \over \pi}\sum_{r=-\infty}^{\infty}
{|\om| \over \om^2 + (p+r
N)^2}
\label{kerp}
\ee
where $\khi^{(p)}_l =
\sqrt{2/N} \sin{\pi p l \over N},\ \  p=1,2,...,N-1$
($\khi^{(1)}$ is the
Perron-Frobenius mode) is the orthogonal set of
eigenfunctions of the Cartan matrix $C_{ln} = 2 \delta_{ln}
-\delta_{l,n+1}- \delta_{l+1,n}$.

Let us note that the expression (\ref{kerp}) for the kernel of the
integral equation reminds the propagator of the free motion of a
particle on the space consisting from the physical space (the time
plus the 1-dimensional physical space described by rapidities) plus
the discrete periodized space represented by the Dynkin diagram. In
this sense it looks like the effective space of our model becomes
three dimensional, the phenomenon familiar to us from the experience
with the c=1 bosonic string (where it effectively passes from 1D to 2D).

Then the density $\rho$ obeys the equation:
\be
{1\over N} \int_{-B}^{B}
R^{(1)}(\te-\te^{\prime}) \rho(\te^{\prime}) {d\te^{\prime}} =
 {m \over 2\pi} \cosh \te.
\label{equp}
\ee

Further simplifications occur in the large $N$ limit
\be
R^{(1)}(\om)\approx
{2N \over \pi} {|\om| \over   \om^2 + 1 }
\label{kerN}
\ee
Now the density $\rho$ may be found in a closed form. To see this, let
us apply the operator $(-{\d^2 \over \d \te^2} +1)$ on both sides of
the equation. As a result we obtain an integral equation with the
Cauchy kernel $(\te-\te')^{-2}$. This equation is solvable:
\be
\rho(\te) =  {m\over 4 K_0(B)\sqrt{B^2-\te^2}}
\label{semi}
\ee
where $K_0(B)$ is the Bessel function.  Note that, in the large $N$
limit $R^{(p)}(\om)$ vanishes at large $\om$, whereas at finite $N$ it
approaches $1$ (see eq.(28)). This implies a singular behaviour of
$\rho(\te)$ at the Fermi point $\pm B$. As a result the physics on the
threshold $h\sim m$ will be changed drastically.

The value of the Fermi rapidities
 as a function of number of particles  can now be
obtained from
\be
 n=\int_{-B}^{B}\rho(\te) d\te ={\pi m \over 4 K_0(B)}
\label{n}
\ee
In  turn the  energy of the state with a given number of particles is
\be
{ E/N^2}={m\over 2\pi^2 }
\int _{-B}^{B} \cosh\te\rho(\te)d\te={m^2\over 8\pi}{I_0(B)\over K_0(B)}
\label{grst}
\ee
And finally, the field $h=-{2\pi^2/N^2} dE/dn$ which corresponds
to a given number of
particles and the energy as a function of the field ${\cal E}=E-\sum_l
h_l n_l = E - N^2/(2\pi^2) \ h n$ are
given by the formulae (\ref{mh}) and (\ref{fh}).

 Let us note that in the large $N$ limit any virtual and real
processes involve all particles, since minimal energies are greater
then a minimal separation between masses.  A reasonable external field
$h_l\sim m_l$ excites all of them on equal footing and leads to
collective effects.

\section{Conclusions}

Let us now list the main lessons which can be drawn out from the
solution:

1. Renormalized planar graphs do not model the world sheets of any
string (they have the zero radius of convergency, due to renormalons).

2.A possible ``stringy'' behaviour with the 2 dimensional physical
target space  may occur only in the non-perturbative strong
coupling regime (near the threshold ($h \sim m$). In this domain the
physics looks similar to the $c=1$ bosonic string described by the
matrix quantum mechanics \cite{Kaz}. At least, the ground state energy
contains the similar inverse logarithmic behaviour with the scale
paramenter, and the fermionic spectrum of excitations which becomes
classical in the large N limit, is also common for two models.

3. The Bethe Ansatz solution of the PCF can be effectively described
as having (1+1+1) dimensions: (time$\times$space$\times$Dynkin
diagram), according to the form of the integral equations for
densities of particles.
The picture is similar to (time $\otimes$ Liouville (eigenvalue) mode)
forming a (1+1) dimensional effective space in the c=1 bosonic string.

Among the most obvious questions which are left we list the following
ones:

1. Elaboration of the $1/N$ expansion and of the double scaling limit
around the threshold $h \sim m$, where a possibility for the
``stringy'' behaviour exists.

2. One has to understand the role of the 3-rd dimension (Dynkin
diagram) in the model. May be, the theory is ``almost free'' in the
(1+1+1) dimensional effective space.

3. Main challenge: to solve the model by simpler (matrix model)
methods directly in terms of the original chiral field $g(x)$. The
character expansion used for the 1D principal chiral field (equivalent
to the heat kernel on the group $SU(N)$ manifold) \cite{KW,GM} might
be very useful for it.

\medskip

{\large\bf Acknowledgements}

I would like to thank the organizers of the les Houches Summer-94 School
for the excellent organization and a warm and creative atmosphere
during  my stay in Les Houches.


\begin{thebibliography}{99}

\medskip
\bibitem{FKW}
V. A. Fateev, V. A. Kazakov, P. B. Wiegmann,
Nucl. Phys. B424 (1994) 505;
Phys. Rev. Lett. 73 (1994) 1750.
\bibitem{PolWieg}  A. M. Polyakov, P. B. Wiegmann, Phys. Lett. 131B (1984)
121.
\bibitem{W1}
P. B. Wiegmann, Phys. Lett. 141B (1984) 217.
\bibitem{W2} P. B. Wiegmann,
 Phys. Lett. 142B (1984) 173.
\bibitem{Ab}  E. Abdalla, M. Abdalla, A. Lima-Santos,
Phys. Lett. 140B, 71 (1984).
\bibitem{Kaz}  V. A. Kazakov, A. A. Migdal, Nucl. Phys. B311 (1988) 171,
for review see: V. A. Kazakov,  in: "Random surfaces and quantum
gravity", Edited by O. Alvarez et al., Plenum Press, New-York (1991), 269.
\bibitem{KW}
V. A. Kazakov, T. Wynter, preprint LPTENS-94/28 (October 1994).
\bibitem{GM}
D. Gross, A. Matytsin, Princeton preprint PUPT 1503 (October 1994).
\end{thebibliography}
\end{document}